%Paper: gr-qc/9403047
%From: j_halliwell@icva.DNET.NASA.GOV
%Date: Sat, 26 Mar 94 10:20:28 -0500

% This paper runs in plain Tex. It include the necessary parts of
% jnl.tex and reforder.tex.
%
%
\font\twelverm=cmr10 scaled 1200    \font\twelvei=cmmi10 scaled 1200
\font\twelvesy=cmsy10 scaled 1200   \font\twelveex=cmex10 scaled 1200
\font\twelvebf=cmbx10 scaled 1200   \font\twelvesl=cmsl10 scaled 1200
\font\twelvett=cmtt10 scaled 1200   \font\twelveit=cmti10 scaled 1200
\font\twelvesc=cmcsc10 scaled 1200  %\font\twelvesf=cmssmc10 scaled 1200
\skewchar\twelvei='177   \skewchar\twelvesy='60

%  Define \...point macros to change fonts and spacings consistently

\def\twelvepoint{\normalbaselineskip=12.4pt plus 0.1pt minus 0.1pt
  \abovedisplayskip 12.4pt plus 3pt minus 9pt
  \belowdisplayskip 12.4pt plus 3pt minus 9pt
  \abovedisplayshortskip 0pt plus 3pt
  \belowdisplayshortskip 7.2pt plus 3pt minus 4pt
  \smallskipamount=3.6pt plus1.2pt minus1.2pt
  \medskipamount=7.2pt plus2.4pt minus2.4pt
  \bigskipamount=14.4pt plus4.8pt minus4.8pt
  \def\rm{\fam0\twelverm}          \def\it{\fam\itfam\twelveit}%
  \def\sl{\fam\slfam\twelvesl}     \def\bf{\fam\bffam\twelvebf}%
  \def\mit{\fam 1}                 \def\cal{\fam 2}%
  \def\sc{\twelvesc}               \def\tt{\twelvett}
  \def\sf{\twelvesf}
  \textfont0=\twelverm   \scriptfont0=\tenrm   \scriptscriptfont0=\sevenrm
  \textfont1=\twelvei    \scriptfont1=\teni    \scriptscriptfont1=\seveni
  \textfont2=\twelvesy   \scriptfont2=\tensy   \scriptscriptfont2=\sevensy
  \textfont3=\twelveex   \scriptfont3=\twelveex  \scriptscriptfont3=\twelveex
  \textfont\itfam=\twelveit
  \textfont\slfam=\twelvesl
  \textfont\bffam=\twelvebf \scriptfont\bffam=\tenbf
  \scriptscriptfont\bffam=\sevenbf
  \normalbaselines\rm}

%       tenpoint

%%
%%      Various internal macros
%%

\def\beginlinemode{\endmode
  \begingroup\parskip=0pt \obeylines\def\\{\par}\def\endmode{\par\endgroup}}
\def\beginparmode{\endmode
  \begingroup \def\endmode{\par\endgroup}}
\let\endmode=\par
{\obeylines\gdef\
{}}
\def\singlespace{\baselineskip=\normalbaselineskip}

\def\oneandahalfspace{\baselineskip=\normalbaselineskip
  \multiply\baselineskip by 3 \divide\baselineskip by 2}
\def\doublespace{\baselineskip=\normalbaselineskip \multiply\baselineskip by 2}

\newcount\firstpageno
\firstpageno=2
\footline={\ifnum\pageno<\firstpageno{\hfil}\else{\hfil\twelverm\folio\hfil}\fi}
\def\toppageno{\global\footline={\hfil}\global\headline
  ={\ifnum\pageno<\firstpageno{\hfil}\else{\hfil\twelverm\folio\hfil}\fi}}
\let\rawfootnote=\footnote              % We must set the footnote style
\def\footnote#1#2{{\rm\singlespace\parindent=0pt\parskip=0pt
  \rawfootnote{#1}{#2\hfill\vrule height 0pt depth 6pt width 0pt}}}
\def\raggedcenter{\leftskip=4em plus 12em \rightskip=\leftskip
  \parindent=0pt \parfillskip=0pt \spaceskip=.3333em \xspaceskip=.5em
  \pretolerance=9999 \tolerance=9999
  \hyphenpenalty=9999 \exhyphenpenalty=9999 }
\def\dateline{\rightline{\ifcase\month\or
  January\or February\or March\or April\or May\or June\or
  July\or August\or September\or October\or November\or December\fi
  \space\number\year}}
\def\received{\vskip 3pt plus 0.2fill
 \centerline{\sl (Received\space\ifcase\month\or
  January\or February\or March\or April\or May\or June\or
  July\or August\or September\or October\or November\or December\fi
  \qquad, \number\year)}}

%%
%%      Page layout, margins, font and spacing (feel free to change)
%%

\hsize=6.5truein
%\hoffset=1truein
\vsize=9.9truein  %% 8.9in
\voffset=-1.0truein
\parskip=\medskipamount
\def\\{\cr}
\twelvepoint            % selects twelvepoint fonts (cf. \tenpoint)
\doublespace            % selects double spacing for main part of paper (cf.
                        %       \singlespace, \oneandahalfspace)
\overfullrule=0pt       % delete the nasty little black boxes for overfull box

\def\title                      %  Title on title page
  {\null\vskip 3pt plus 0.2fill
   \beginlinemode \doublespace \raggedcenter \bf}

\def\author                     %  Author(s) name(s)  on title page
  {\vskip 3pt plus 0.2fill \beginlinemode
   \singlespace \raggedcenter\sc}

\def\affil                      % Affiliations (can intermix with \author)
  {\vskip 3pt plus 0.1fill \beginlinemode
   \oneandahalfspace \raggedcenter \sl}

\def\abstract                   % Begin abstract
  {\vskip 3pt plus 0.3fill \beginparmode
   \singlespace ABSTRACT: }

\def\endtopmatter               % End title page, begin body of paper
  {\endpage                     %       This subsumes \body
   \body}

\def\body                       % Begin text body;  can be used to end
  {\beginparmode}               % \title, \author, \affil, \abstract,
                                % \reference, or \figurecaption modes

\def\head#1{                    % Head;  NOTE enclose the text in {}
  \goodbreak\vskip 0.5truein    %  e.g., \head{I. Introduction}
  {\immediate\write16{#1}
   \raggedcenter \uppercase{#1}\par}
   \nobreak\vskip 0.25truein\nobreak}

\def\beginitems{
\par\medskip\bgroup\def\i##1 {\item{##1}}\def\ii##1 {\itemitem{##1}}
\leftskip=36pt\parskip=0pt}
\def\enditems{\par\egroup}

\def\beneathrel#1\under#2{\mathrel{\mathop{#2}\limits_{#1}}}

\def\refto#1{$^{#1}$}           % For references in text as superscript

\def\references                 % Begin references -- basic format is Phys Rev
  {\head{References}            % I.e., volume, page, year (space after
%%commas).
   \beginparmode
   \frenchspacing \parindent=0pt \leftskip=1truecm
   \parskip=8pt plus 3pt \everypar{\hangindent=\parindent}}

\gdef\refis#1{\item{#1.\ }}                     % Ref list numbers.

\gdef\journal#1, #2, #3, 1#4#5#6{               % Journal reference.  Comma
%%sets
    {\sl #1~}{\bf #2}, #3 (1#4#5#6)}            % off: name, vol, page, year

\gdef\refa#1, #2, #3, #4, 1#5#6#7.{\noindent#1, #2 {\bf #3}, #4 (1#5#6#7).\rm}
%refa: type in: name,
%journal, vol, page, year
%prints out in same order

\gdef\refb#1, #2, #3, #4, 1#5#6#7.{\noindent#1 (1#5#6#7), #2 {\bf #3}, #4.\rm}
%refb: reads in same
%prints out name (year) etc.

\def\endreferences{\body}

\def\endpage                    %  Eject a page
  {\vfill\eject}

\def\endpaper                   %  Ways to say goodbye
  {\endmode\vfill\supereject}

\def\ref#1{Ref.~#1}                     %       for inline references
\def\Ref#1{Ref.~#1}                     %       ditto
\def\[#1]{[\cite{#1}]}
\def\cite#1{{#1}}
%%\def\Equation#1{Equation~(#1)}                % For citation of equation
%%numbe
%%\def\Equations#1{Equations~(#1)}      %       ditto
%%\def\Eq#1{Eq.~(#1)}                   %       ditto
%%\def\Eqs#1{Eqs.~(#1)}                 %       ditto
\def\(#1){(\call{#1})}
\def\call#1{{#1}}
\def\taghead#1{}
\def\frac#1#2{{#1 \over #2}}
\def\half{{\frac 12}}

\def\12{{1\over2}}

\catcode`@=11
\newcount\r@fcount \r@fcount=0
\newcount\r@fcurr
\immediate\newwrite\reffile
\newif\ifr@ffile\r@ffilefalse
\def\w@rnwrite#1{\ifr@ffile\immediate\write\reffile{#1}\fi\message{#1}}

\def\writer@f#1>>{}
\def\referencefile{%			  Stuff to write .REF file
  \r@ffiletrue\immediate\openout\reffile=\jobname.ref%
  \def\writer@f##1>>{\ifr@ffile\immediate\write\reffile%
    {\noexpand\refis{##1} = \csname r@fnum##1\endcsname = %
     \expandafter\expandafter\expandafter\strip@t\expandafter%
     \meaning\csname r@ftext\csname r@fnum##1\endcsname\endcsname}\fi}%
  \def\strip@t##1>>{}}

\def\citeall#1{\xdef#1##1{#1{\noexpand\cite{##1}}}}
\def\cite#1{\each@rg\citer@nge{#1}}	% Variable No. of args, separated by

\def\each@rg#1#2{{\let\thecsname=#1\expandafter\first@rg#2,\end,}}
\def\first@rg#1,{\thecsname{#1}\apply@rg}	% each@ag is a general purpose
\def\apply@rg#1,{\ifx\end#1\let\next=\relax%	  variable no. of arg. macro.
\else,\thecsname{#1}\let\next=\apply@rg\fi\next}% args separated by commas

\def\citer@nge#1{\citedor@nge#1-\end-}	% Check for M-N range (M and N numbers)
\def\citer@ngeat#1\end-{#1}
\def\citedor@nge#1-#2-{\ifx\end#2\r@featspace#1 % Single argument
  \else\citel@@p{#1}{#2}\citer@ngeat\fi}	% M-N range of arguments
\def\citel@@p#1#2{\ifnum#1>#2{\errmessage{Reference range #1-#2\space is bad.}%
    \errhelp{If you cite a series of references by the notation M-N, then M and
    N must be integers, and N must be greater than or equal to M.}}\else%
 {\count0=#1\count1=#2\advance\count1 by1\relax\expandafter\r@fcite\the\count0,
  \loop\advance\count0 by1\relax%	  Loop from M to N
    \ifnum\count0<\count1,\expandafter\r@fcite\the\count0,%
  \repeat}\fi}

\def\r@featspace#1#2 {\r@fcite#1#2,}	% Eat spaces at beginning or end of arg
\def\r@fcite#1,{\ifuncit@d{#1}%		  Cite individual reference
    \newr@f{#1}%
    \expandafter\gdef\csname r@ftext\number\r@fcount\endcsname%
                     {\message{Reference #1 to be supplied.}%
                      \writer@f#1>>#1 to be supplied.\par}%
 \fi%
 \csname r@fnum#1\endcsname}
\def\ifuncit@d#1{\expandafter\ifx\csname r@fnum#1\endcsname\relax}%
\def\newr@f#1{\global\advance\r@fcount by1%
    \expandafter\xdef\csname r@fnum#1\endcsname{\number\r@fcount}}

\let\r@fis=\refis			% Save old \refis, redefine
\def\refis#1#2#3\par{\ifuncit@d{#1}%      Use two params #2 #3 to strip blank
   \newr@f{#1}%
   \w@rnwrite{Reference #1=\number\r@fcount\space is not cited up to now.}\fi%
  \expandafter\gdef\csname r@ftext\csname r@fnum#1\endcsname\endcsname%
  {\writer@f#1>>#2#3\par}}

\def\ignoreuncited{%   redefine \refis if ignoring uncited references
   \def\refis##1##2##3\par{\ifuncit@d{##1}%
    \else\expandafter\gdef\csname r@ftext\csname r@fnum##1\endcsname\endcsname%
     {\writer@f##1>>##2##3\par}\fi}}

\def\r@ferr{\endreferences\errmessage{I was expecting to see
\noexpand\endreferences before now;  I have inserted it here.}}
\let\r@ferences=\references
\def\references{\r@ferences\def\endmode{\r@ferr\par\endgroup}}

\let\endr@ferences=\endreferences
\def\endreferences{\r@fcurr=0%		  Save old \endreferences, redefine
  {\loop\ifnum\r@fcurr<\r@fcount%	  Loop over refnum and produce text
    \advance\r@fcurr by 1\relax\expandafter\r@fis\expandafter{\number\r@fcurr}%
    \csname r@ftext\number\r@fcurr\endcsname%
  \repeat}\gdef\r@ferr{}\endr@ferences}

% Save old \endpaper, redefine it to write parting message.

\let\r@fend=\endpaper\gdef\endpaper{\ifr@ffile
\immediate\write16{Cross References written on []\jobname.REF.}\fi\r@fend}

\catcode`@=12

\citeall\refto		% These macros will generate citations
\citeall\ref		%
\citeall\Ref		%

\def\a{{\alpha}}

\def\half{{1 \over 2}}
\def\ra{{\rangle}}
\def\la{{\langle}}
\def\dt{{\delta t}}
\def\Li{{L_I}}
\def\Lr{{L_R}}
\def\ih{{i \over \hbar}}
\def\l{\ell}
\def\E{{\cal E}}
\def\au{\underline \alpha}

\centerline{\bf Decoherent Histories and Quantum State Diffusion}

\vskip 0.3in
\author Lajos Di\'osi%\footnote{$^{\dag}$}{E-mail address: ???}
\affil
KFKI Research Institute for Particle and Nuclear Physics
H-1525 Budapest 114, POB 49
HUNGARY
\bigskip
\author Nicolas Gisin
\affil
Group of Applied Physics, University of Geneva
1211 Geneva 4
SWITZERLAND
\bigskip
\author Jonathan Halliwell%\footnote{$^{*}$}{E-mail
%address:j\_halliwell@vax1.physics.imperial.ac.uk}
\affil
Theory Group, Blackett Laboratory
Imperial College, London SW7 2BZ
UK
\vskip 0.2in
\centerline {and}
\vskip 0.1in
\author Ian C. Percival
\affil
Department of Physics, Queen Mary and Westfield College
Mile End Road, London E1 4NS
UK
\vskip 0.5in
\centerline{\rm PACS Numbers: 03.65.-w, 03.65.Bz, 05.40.+j, 42.50.-p}
\vskip 0.2in
\centerline {\rm Preprint IC 93-94/25. March, 1994}
\vskip 0.2in
\centerline {\rm Submitted to {\sl Physical Review Letters}}

\abstract
{We demonstrate a close connection between the
decoherent histories (DH) approach to quantum mechanics and the
quantum state diffusion (QSD) picture, for open
quantum systems described by a master equation of Lindblad form.
The (physically unique) set of
variables that localize in the QSD picture also
define an approximately decoherent set of histories in the DH approach.
The degree of localization is related to the degree of decoherence,
and the probabilities for histories prescribed by each approach
are essentially the same. }
\endtopmatter

The last decade has witnessed considerable interest in
the foundations of quantum mechanics. Many reasons for this may be
found: the long-felt dissatisfaction with the
Copenhagen interpretation; certain experimental developments and the
Copenhagen interpretation's inability to supply a useful qualitative
account of them; the special needs of quantum mechanics applied to
the entire universe (quantum cosmology); and the general desire
to possess a deeper understanding of quantum theory.
Partially for
these reasons, modifications and generalizations of both the mathematics
and interpretation of quantum theory have been sought.

This letter is concerned with demonstrating the connections between two
recently developed alternative approaches to quantum theory, each of
which was proposed, independently, with the aim of shedding light
on some of the difficulties outlined above. The approaches we shall
compare are the decoherent (or ``consistent'') histories approach
[\cite{HG,DH}] and the quantum state diffusion picture
[\cite{GP1,GP2,GP3}]
(see Ref.[\cite{DioDH}] for an early guess at the relation between
these two approaches).
In both of these approaches, the basic
mathematical formalism of quantum mechanics is left untouched, but
new insight into its interpretation is obtained by focusing on
different types of mathematical objects.

We will be concerned with
a quantum system consisting of a subsystem coupled to
its environment. The subsystem is then frequently
referred to as an open quantum system, and we shall do so here.
Conventionally, an open system is
described by a reduced density operator $\rho$, evolving according
to a master equation, derived by tracing over the environment.
Under the assumption that the evolution is
Markovian, the master equation takes the Lindblad form [\cite{Lind}],
$$
{ d \rho \over dt} = -\ih  [H, \rho] - \half \sum_{j=1}^n \left(
\{ L_j^{\dag} L_j, \rho \} - 2 L_j \rho L_j^{\dag} \right)
\eqno(1)
$$
Here, $H$ is the Hamiltonian of the open system in the absence of
the environment (sometimes modified by terms depending on the $L_j$)
and the $n$ operators $L_j$ model the effects of the environment.
This equation is frequently used in quantum optics [\cite{Carm}] and in
studies of decoherence [\cite{Zur,Zeh}].
For example, in the much-studied quantum
Brownian motion model [\cite{Cald,DioCL,Sal}],
the master equation is (1) with
a single Lindblad operator $ L = (2D)^{-\half} ( \hat x +
2 \ih \gamma D \hat p
) $, with $D = \hbar^2 /(8 m \gamma k T) $ (where $\gamma$ is the
dissipation and $T$ is the temperature of the environment),
and $H = H_S + \half \gamma \{ \hat x, \hat p\} $,
where $H_S$ is the distinguished subsystem Hamiltonian, in the absence
of the environment.

Density
operators satisfying (1) give statistical predictions in full
agreement with experiment in a wide variety of situations. However,
they do not give a picture of the behaviour
of an {\it individual} system, but only of ensembles.
The quantum state diffusion approach
avoids this shortcoming [\cite{GP1,GP2,GP3,Perc}]. It originated from
considerations of the quantum measurement problem
[\cite{Pearle,GiQSD,DioQSD, DioIto}],
and the desiderata of describing individual experimental outcomes and
putting physical intuition into the equations, as urged by Bell [\cite{Bell}].
It was also motivated by its computational advantage and insight in
treating practical problems in open systems [\cite{Perc2}].
It consists of an ``unravelling'' of the
evolution of $\rho$ under (1). This involves
regarding $\rho$ as a mean over a distribution of pure state
density operators, $ \rho = M | \psi \ra \la \psi | $,
where $M$ denotes the mean (defined below),
with the pure states evolving according to
the non-linear stochastic Langevin-Ito equation,
$$
\eqalignno{
| d \psi \ra = -\ih H |\psi \ra dt & + \half \sum_j \left(
2 \la L_j^{\dag} \ra L_j - L_j^{\dag} L_j - \la L_j^{\dag} \ra
\la L_j \ra \right) | \psi \ra \ dt
\cr &
+ \sum_j \left( L_j - \la L_j \ra \right) | \psi \ra \ d \xi_j(t)
&(2) \cr }
$$
for the normalized state vector $| \psi \ra $.
Here, the $d \xi_j$ are independent complex differential random
variables representing a complex Wiener process. Their linear and
quadratic means are,
$ M [ d \xi_j d \xi_k^* ] = \delta_{jk} \ dt $,
$ M[ d \xi_j d \xi_k ] = 0 $ and $ M [ d\xi_j ] = 0 $.

The quantum state diffusion picture described by the Ito equation (2)
is mathematically equivalent to the Lindblad equation (1). However, its
appeal lies in the fact that the
solutions to the Ito equation appear to correspond rather well to
individual experimental runs, and thus provide considerable insight
into the behaviour of individual processes and systems.
Solutions to the Ito equation commonly have the
property of {\it localization} -- the dispersion of certain
operators tends to decrease as time evolves. This has been
demonstrated by numerical solutions [\cite{GP3}], analytic solutions in
special cases [\cite{Sal,DioIto,DioAnalLoc}],
and some general theorems [\cite{GP2,Perc}].
The method has also been
successfully used to analyze quantum jump experiments
[\cite{GKThW}].

Our first results, required for the comparison with the decoherent
histories approach below, are explicit representations of the
solutions to the Lindblad equation (1) and the Ito equation (2).
To solve the Lindblad equation, consider the case of a single
Lindblad operator $L = L_R + i L_I $, where $L_R$, $L_I$ are hermitian.
Divide the finite time interval $[0,t]$
into $K$ subintervals, so that $ t = K \dt $, and let $\dt \rightarrow
0$, $ K \rightarrow \infty $, holding $t$ constant.
Then we have the following representation
solution to (1):
$$
\eqalignno{
\rho(t ) & = { {\rm lim} \atop \dt \rightarrow 0, K \rightarrow \infty}
\ \left( { \dt \over \pi } \right)^K
\int  d^2\l_1 \cdots d^2\l_K
\cr & \times
\prod_{m=1}^K
\ \exp \left( {\dt \over 2} ( \ell^*_m L - \ell_m L^{\dag} ) \right)
\ \exp \left( - {\dt \over 2 } \ |L -\l_m|^2 \right)
\ \exp \left( - \ih H_0 \dt \right)
\rho(0) \cr
& \times
\prod_{m=1}^K
\ \exp \left( \ih H_0 \dt \right)
\ \exp \left( - { \dt \over 2 } \ |L  -\l_m|^2 \right)
\ \exp \left( - { \dt \over 2 } (\ell^*_m L - \ell_m L^{\dag} )\right)
&(3) \cr }
$$
where $H_0 = H + {i \hbar \over 4} [L, L^{\dag}]$,
and the $\ell_m$ are complex numbers at the discrete moments of time
labeled by $m$.
We use the notation
$|L-\l_m|^2\equiv
\left(\Lr-{\rm Re} \l_m\right)^2+\left(\Li-{\rm Im}\l_m\right)^2$.
That this is the solution is
readily verified by explicit computation [\cite{DGHP}].
The solution has the form
of a ``measurement process" of the $L$'s, continuous in time,
with ``feedback" via the terms $ (\ell^*_m L - \ell_m L^{\dag} )$
[\cite{Caves}]. The case of many
different Lindblad operators, $L_j$, is readily obtained by taking
products over $j$ of the appropriate operators in (3) at each moment of time.
The ordering of the
operators at each moment of time
is irrelevant in the limit $\dt \rightarrow 0 $
(although the operators at different times are time-ordered,
according to increasing $m$).
For future reference we write Eq.(3) in terms of a density operator
propagator as $\rho(t) = K_0^t [\rho(0)]$.

Similarly, the solution to the Ito equation has the explicit
representation
$$
\eqalignno{
|\psi (t)\ra =\lim_{\delta t \rightarrow 0} \prod_{m=1}^K
& \exp\left( \frac{\delta t}{2}
\left(
2 \la L^{\dag} \ra_m L
- L^{\dag} L - \la L^{\dag} \ra_m \la L \ra_m
\right)
+ ( L - \la L \ra_m ) \delta \xi \right) \cr
\times
& \exp \left( - \ih H \delta t \right) |\psi (0)\ra.
&(4) \cr }
$$
where $\la L \ra_m $ denotes the expectation of $L$ at time
$t_m = m \dt $.
Eq.(4) expresses an individual history $|\psi_{\xi}(t)\ra$ as an
explicit functional of
an individual complex gaussian noise $\xi(t)$.
However, to make clearer the connection with Eq.(3) (and ultimately
with the decoherent histories approach discussed below),
it is useful to present an
alternative representation, obtained by exchanging the Wiener
process $\xi(t)$ for the stochastic variable $\l(t)$, defined by
$( \ell(t) - \la L \ra ) dt = d \xi^*(t) $ (Ref.[\cite{DioIto}]).
Eq.(4) thus becomes
$$
\eqalignno{
| \psi(t) \rangle & = { {\rm lim} \atop \dt \rightarrow 0, K
\rightarrow \infty}
\left( { \dt \over \pi } \right)^{K/2} \ N (\ell_1 \cdots \ell_K)
\cr  \times
\  & \prod_{m=1}^K
\ \exp \left( {\dt \over 2} ( \ell^*_m L - \ell_m L^{\dag} ) \right)
\ \exp \left( - { \dt \over 2} \ |L -\l_m|^2 \right)
\ \exp \left( - \ih H_0 \dt \right)
| \psi(0) \rangle
&(5) \cr }
$$
where $H_0$ is given above after Eq.(3).
$N$ is a (complex) normalization factor, ensuring that $ \la \psi (t) |
\psi(t) \ra =1 $, and appears because the state is now to be
thought of as a functional of the stochastic process $\l(t)$ instead
of $\xi(t)$, $| \psi \ra = | \psi_{\l} \ra $.
Again the generalization to many Lindblad generators is straighforward.
That (4), (5) are solutions to Eq.(2)
may be verified by explicit computation [\cite{DGHP}].
This explicit representation of the solution clearly indicates
that the solutions will tend to exhibit localization in
the operator $L$. It also illustrates that the solutions have the
form of a ``trajectory'', or ``history'',
concentrated around the $\l(t)$ corresponding
to the particular Wiener process $\xi(t)$.

We may associate a probability with these trajectories.
To see this, recall that we are working from the outset
with a statistical ensemble of
solutions to the Ito equation, $|\psi_{\xi} \ra $, obeying the
rules of standard (``classical'') probability theory, and the
probability distribution of the solutions is that implied by the
means of $d \xi$, {\it etc.}, given above. Indeed, the solution (3) to
the master equation, is a sum over $\xi$ of $| \psi_{\xi} \ra \la
\psi_{\xi} |$, weighted by the probability for each solution.
As before, it is more useful to express this probability
distribution as a probability $p[\l(t)]$ over the states $|\psi_\l
\ra$ satisfying Eq.(5). Since Eq.(3) must be a mean over
$ | \psi_{\l} \ra \la \psi_{\l} |$, it is easily seen from Eqs.(3)
and (5), that the probability distribution over the
$|\psi_{\l}\ra$'s must be $ p [ \l(t) ] = \bigl| N [\l(t)] \bigr|^{-2} $
(or what amounts to the same, the norm of the state (5) but without the
normalization factor $N$). Note, however, that in general
$\la \psi_{\l'} | \psi_{\l} \ra \ne 0$
for $\l \ne \l'$, and thus $p[\l(t)]$ may be
thought of as a probability for {\it histories of values of}
$L$ only when the solutions $|\psi_{\l} \ra $  are
reasonably well-localized in $L$.

The Lindblad and Ito equations, (1), (2), are invariant under
redefinitions of the $L_j$'s by unitary transformations, {\it i.e.},
$ L_j \rightarrow L'_j = \sum_{k=1}^n U_{jk} L_k $, where $U_{kj}$ is a
unitary matrix. For the case $n=1$, this is just multiplication by a
constant phase, and the solutions (3), (4), do indeed manifest
this invariance [\cite{GP1}].

Turn now to the decoherent histories approach.
The decoherent histories approach is a generalization of quantum
mechanics to genuinely closed systems, such as the entire  universe
[\cite{HG,DH}].
Its aim is to give a predictive formulation of quantum theory
applicable to closed systems which does not rely on notions of measurement
or on the existence of an external classical domain. From such a
framework, one hopes to understand the emergence of the
classical world from an underlying quantum one, and the origin of
the quantum-classical division upon which the Copenhagen
interpretation depends.

In the decoherent histories approach, the mathematical objects
one focuses on are the probabilities for histories of a
closed system. A quantum-mechanical history is defined by an initial
state $\rho$ at time $t=0$ together with a string of projection operators
$ P_{\a_1} \cdots P_{\a_n}$ acting at times $t_1 \dots t_n$,
characterizing the possible alternatives of the system at those times.
The projections are exhaustive, $\sum_{\a} P_{\a} =1 $, and exclusive,
$ P_{\a} P_{\beta} = \delta_{\a \beta} P_{\a}$.
Because
of interference, most sets of histories for a closed system cannot be
assigned probabilities. The interference between pairs of histories
in a set is measured by the so-called decoherence functional,
$$
D(\au, \au' ) = {\rm Tr} \left( P_{\a_n}(t_n) \cdots P_{\a_1}(t_1)
\rho P_{\a_1'}(t_1)  \cdots P_{\a_n'} (t_n) \right)
\eqno(6)
$$
where $P_{\a_k} (t_k) = e^{-\ih H t_k } P_{\a} e^{\ih H t_k} $,
$H$ is the Hamiltonian of the closed system
and $\au$ denotes the string $\a_1 \cdots \a_n $.
When $D(\au,\au') \approx 0 $
for $ \au \ne \au'$, inteference may be neglected, and one may
assign the probability $p(\au) = D(\au, \au)$ to the history.
Probabilities assigned under this condition may be shown to
obey the sum rules of probability theory [\cite{HG}].
Sets of histories satisfying this condition are called decoherent.
Loosely, satisfaction of these conditions means that one can
``talk about'' ({\it i.e.}, apply classical logic to) the physical
properties of the system, and think about those properties as if
they were definite, without having to invoke notions of measurement.
Given the Hamiltonian and initial state for a closed system, one's
initial aim is to determine the strings of projection operators for
which the decoherence condition is satisfied.

The decoherent histories approach is readily applied to the class of
open systems considered here, {\it i.e.}, closed systems in which
there is a natural separation into a distinguished subsystem and the
rest.   For such systems, a natural set of
histories to study are those characterized by the properties of the
distinguished subsystem at each moment of time, but ignoring ({\it
i.e.}, coarse-graining over) the properties of the environment.
Histories of this type are often decoherent as a result of the
interaction between the system and the environment.  To be precise,
consider histories characterized by strings of projections $P_{\a_1}
\otimes I^{\E} \cdots P_{\a_n} \otimes I^{\E} $  at times $t_1
\cdots t_n$, where $I^{\E}$ denotes the identity on the environment.
Now assuming that the initial density operator factorizes, the
trace over the environment may be carried  out explicitly in the
decoherence functional (6), and, in the regime in which a Markovian
approximation holds, it then has the form
$$
D(\au, \au') =
{\rm Tr} \left(
P_{\a_n} K_{t_{n-1}}^{t_n}[P_{\a_{n-1}}\cdots
K_{t_1}^{t_2}[P_{\a_1} K_0^{t_1}[\rho(0)] P_{\a'_1}]
\cdots P_{\a'_{n-1}}] \right)
\eqno(7)
$$
Here,
$K_{t_k}^{t_{k+1}} $ is the reduced density operator
propagator introduced above, and
the trace is now over the distinguished subsystem only.

Given this expression, and given the explicit form of $K$ above,
Eq.(3), we may now discuss decoherence. For simplicity, consider the
case of projections continuous in time in the decoherence functional (7).
The discrete time version of the decoherence functional will
contain terms of the form,
$$
P_{\a_k} \ K_{t_{k-1}}^{t_k} [ \cdots ] \ P_{\a_k'}
= \int d^2 \ell  \ P_{\a_k} \ \exp \left( - { \dt \over 2}
| L - \l|^2 \right)[ \cdots
] \ \exp \left( - { \dt \over 2} |L - \l|^2 \right) \ P_{\a_k'}
\eqno(8)
$$
The operator $\exp \left( -{ \dt \over 2 } |L-\l|^2 \right) $
is an approximate projection operator in the limit, used here,
of small $\dt$.
(It is an approximate projector for all $\dt$ if $ [L,L^{\dag}]=0$).
Now the key point is that on the right hand side,
we have two different projection operators
$P_{\a_k}$, $P_{\a_k'}$ operating on {\it the same}
Gaussian projection, $\exp \left( -{ \dt\over 2} |L-\l|^2 \right) $.
Because Gaussian projections are approximately exclusive,
the decoherence functional will be
approximately diagonal in the $\a_k$'s if we choose the
projections $P_{\a_k}$ also to be Gaussian projections onto $L$:
$$
P_{\a} = \exp \left( - \half \kappa^2 \dt
\bigl|L - \kappa^{-1} \dt^{-\half} \a  \bigr|^2
\right)
\eqno(9)
$$
Here, $\alpha$ is a dimensionless and {\it complex} continuous label,
and $\kappa^{-1}(\dt)^{-\half}$ is the width of the projection which will be
tuned by the dimensionless parameter $\kappa$. The approximate
exclusivity of these approximate projectors means that $\a$, although
continuous, has significance only up to order $1$. Clearly,
with the choice (9), (8) will be very small unless
$\a\approx\kappa \dt^{\half}\l\approx\a'$.
Therefore, histories characterized by strings of
projections onto $L$ will approximately decohere. We thus arrive
at our main result: the variables exhibiting localization in
the quantum state diffusion picture are the same as the variables
characterizing a decoherent set of histories in the decoherent
histories approach.

The operator (9) is an approximate projection operator (under the
conditions stated above) onto a subset
of the spectrum of the (generally non-hermitian) operator $L$,
and the label $\a$ is complex.
Projections of precisely this type have not previously been used in
the decoherent histories approach, but there is no obvious
obstruction to doing so. Indeed, the use of such projections is
strongly suggested by form of Eq.(8), which arises as a result of
the invariance of Eq.(3) under $ L \rightarrow e^{i \phi} L$.
Furthermore, the connection with
histories characterized by the more familiar types of projectors may
be made by specialization to the case of a hermitian $L$ (strictly,
to the class of operators equivalent to a hermitian operator under
multiplication by a phase).

The degrees of localization and decoherence also are related. From
the solution to the Ito equation (5), the degree of localization is
determined by the degree to which $L$ becomes concentrated about a
particular trajectory $\l(t)$.  At each time step $\dt$, the
localization width
of $L$ is of order $\dt^{-\half}$.  Similarly, from equation (8),
one may see that the degree of decoherence is also determined by the
degree to which $L$ is concentrated about a particular value.
Loosely, the off-diagonal terms of the decoherence functional are
suppressed in comparison to the on-diagonal terms (this is the
appropriate way to measure approximate decoherence [\cite{DH}]) by a
factor of order of  $\exp \left( - \kappa^{-2} |\a-\a'|^2 \times
constant \right) $,  where the constant  is of order 1. This means
that the projectors given by Eq.(9) define an approximately
decoherent set of histories only if $\kappa <<  1$.

The degrees of localization and decoherence are related,
therefore, in the sense that approximate decoherence of histories
may be achieved only if the projectors $P_{\a}$ characterizing the
histories are coarser than the localization width.

Finally, consider the probabilities for histories. Given approximate
decoherence, the decoherent histories approach assigns probabilities
to histories given by the diagonal elements of the decoherence
functional (7).
Consider Eq.(8), but now with
$\a_k = \a_k'$. Under the conditions yielding approximate
decoherence, the integrand in (8) will be very small unless $\l \approx
\kappa^{-1}(\dt)^{-\half} \a$. The projection operators $P_{\a_k}$
then have essentially no effect (except to produce a negligible
modification of the width of the neighbouring Gaussian projectors)
so we can drop them. One thus finds that
$$
P_{\a_k} \ K_{t_{k-1}}^{t_k} [ \cdots ] \ P_{\a_k}
\ \approx   \ \exp \left( - { \dt \over 2 }
| L - \kappa^{-1} \dt^{-\half}
\a_k |^2 \right)[ \cdots
] \ \exp \left( - { \dt \over 2 } |L - \kappa^{-1} \dt^{-\half}
\a_k |^2 \right)
\eqno(10)
$$
Using this result, one may then
see that the probabilities assigned to these histories in the
decoherent histories approach have the form of the norm of the state
(5) without the normalization factor, and with
$\l(t_k) = \kappa^{-1}\dt^{-\half}\a(t_k)$.
They are therefore exactly the same as the probabilities assigned in
the quantum state diffusion approach.
A more detailed account of this work
will be presented elsewhere [\cite{DGHP}]

\noindent{\bf Acknowledgements:} L.D. thanks support by the grant
OTKA No. 1822/1991. J.J.H. was supported by the Royal Society.

\references

\refis{DioCL} L.Di\'osi, {\sl Physica} {\bf A199}, 517 (1993).

\refis{DioDH} L.Di\'osi, {\sl Phys.Lett.} {\bf B280}, 71 (1992).

\refis{DGHP} L.Di\'osi, N.Gisin, J.Halliwell and I.Percival, in preparation.

\refis{HG} M.Gell-Mann and J.B.Hartle, {\sl Phys.Rev.} {\bf D47}, 3345
(1993); R.Griffiths, {\sl J.Stat.Phys.} {\bf 36}, 219 (1984);
R.Omn\`es, {\sl Rev.Mod.Phys.} {\bf 64}, 339 (1992).

\refis{DH} H.F.Dowker and J.J.Halliwell, {\sl Phys.Rev.} {\bf D46}, 1580
(1992)

\refis{GP1} N. Gisin and I.C. Percival, {\sl J.Phys.} {\bf A25},
5677 (1992); see also {\sl Phys. Lett.} {\bf A167}, 315 (1992).

\refis{GP2} N. Gisin and I.C.Percival, {\sl J.Phys.} {\bf A26},
2233 (1993).

\refis{GP3} N. Gisin and I.C.  Percival,  {\sl J.Phys.}
{\bf A26}, 2245 (1993).

\refis{GKThW} N. Gisin, P.L. Knight, I.C. Percival, R.C. Thompson
and  D.C. Wilson, {\sl J. Mod. Optics}, {\bf 40}, 1663 (1993);
B.Garraway and P.Knight, {\sl Phys.Rev.} {\bf A49},
1266 (1994);
P.Goetsch and R.Graham, {\sl Ann.Physik} {\bf 2}, 706 (1993).

\refis{Carm} H.J.Carmichael, {\it An Open Systems Approach to Quantum
Optics} (Springer-Verlag, 1993).

\refis{Pearle} P.Pearle, {\sl Phys.Rev.} {\bf D13}, 857 (1976).

\refis{GiQSD} N.Gisin,
{\sl Phys.Rev.Lett.} {\bf 52}, 1657 (1984); {\sl Hevl.Phys.Acta}
{\bf 62}, 363 (1989).

\refis{DioIto} L.Di\'osi, {\sl Phys.Lett.} {\bf 129A}, 419 (1988).

\refis{DioAnalLoc} L.Di\'osi, {\sl Phys.Lett} {\bf 132A}, 233 (1988).

\refis{Zur} W.H. Zurek, {\sl Physics Today} {\bf 40}, 36 (1991);
J.P.Paz, S.Habib and W.H.Zurek,
\break
{\sl Phys.Rev.} {\bf D47}, 488 (1993).

\refis{Zeh}  H.D. Zeh, {\sl Phys. Lett.} {\bf A172}, 189 (1993);
E.Joos and H.D.Zeh, {\sl Zeit.Phys.} {\bf B59}, 223 (1985).
See
also N. Gisin and I. Percival, {\sl Phys. Lett.} {\bf 175A}, 144 (1993).

\refis{Cald} A.O.Caldeira and A.J.Leggett, {\sl Physica} {\bf 121A},
587 (1983).

\refis{Caves} C.M.Caves and G.J.Milburn, {\sl Phys.Rev.} {\bf A36}, 5543
(1987).

\refis{Lind} G.Lindblad, {\sl Comm.Math.Phys.} {\bf 48}, 119 (1976).

\refis{Perc} I.C.Percival,  {\sl J.Phys.} {\bf A27}, 1003 (1994).

\refis{Perc2} I.Percival, in {\it Quantum Chaos and Quantum
Measurement},
NATO Advanced Study Institute Series, Vol 357, edited by
P.Cvitanovic {\it et al.} (Kluwer, Dordrecht,1992), p.199.

\refis{Bell} J.S.Bell, {\sl Physics World} {\bf 3}, 33 (1990).

\refis{DioQSD} L.Di\'osi, {\sl J.Phys.} {\bf A21}, 2885 (1988)

\refis{Sal} Y.Salama and N.Gisin, {\sl Phys. Lett.}
{\bf 181A}, 269, (1993).

\endreferences

\end